\documentclass[onecolumn,aps,prd,preprintnumbers,showpacs,superscriptaddress,nofootinbib,amsmath,amssymb,floats,floatfix,showkeys,notitlepage,longbibliography]{revtex4-1}

\usepackage{lipsum}
\usepackage{graphicx}
\usepackage{subfigure}
\usepackage{palatino}
\usepackage{changes}
\usepackage{hyperref}
\hypersetup{colorlinks=true,linkcolor=blue,urlcolor=blue,citecolor=blue}
\usepackage[toc,page]{appendix}
\usepackage[normalem]{ulem}
\usepackage{adjustbox}
\usepackage{latexsym}
\usepackage{amsmath}
\usepackage{amssymb}
\usepackage{amsfonts}
\usepackage{times}
\usepackage{dcolumn}
\usepackage{bm}
\usepackage{tikz}
\usepackage{bigints}
\usepackage{array,tabularx,multirow}
\usepackage[tracking=true]{microtype}
\usepackage{color}
\SetTracking{}{500}
\SetTracking{encoding={*}, shape=sc}{40}
\UseRawInputEncoding 
\allowdisplaybreaks

\def\0{{\sst{(0)}}}
\def\1{{\sst{(1)}}}
\def\2{{\sst{(2)}}}
\def\3{{\sst{(3)}}}
\def\4{{\sst{(4)}}}
\def\5{{\sst{(5)}}}
\def\6{{\sst{(6)}}}
\def\7{{\sst{(7)}}}
\def\8{{\sst{(8)}}}
\def\sst#1{{\scriptscriptstyle #1}}

\begin{document} \sloppy

\title{Thermodynamics and logarithmic corrections of symmergent black holes}

\author{Riasat Ali}
\email{riasatyasin@gmail.com}
\affiliation{Department of Mathematics, GC
University Faisalabad Layyah Campus, Layyah-31200, Pakistan}

\author{Rimsha Babar}
\email{rimsha.babar10@gmail.com}
\affiliation{Division of Science and Technology, University of Education, Township, Lahore 54590, Pakistan}

\author{Zunaira Akhtar}
\email{zakhtarmathematics@gmail.com}
\affiliation{Division of Science and Technology, University of Education, Township, Lahore 54590, Pakistan}

\author{Ali \"Ovg\"un}
\email{ali.ovgun@emu.edu.tr}
\homepage{https://www.aovgun.com}
\affiliation{Physics Department, Eastern Mediterranean University, Famagusta, 99628 North Cyprus via Mersin 10, Turkey}

\begin{abstract}
In this paper, we study quantum gravity effect on the symmergent black hole which is derived from quadratic-curvature gravity. To do so, we use the Klein-Gordon equation which is modified by generalized uncertainty principle (GUP). After solving the field equations, we examine the symmergent black hole's tunneling and Hawking temperature. We explore the graphs of the temperature through the outer horizon to check the GUP influenced conditions of symmergent black hole stability. We also explain how symmergent black holes behave physically when influenced by quantum gravity. The impacts of thermal fluctuations on the thermodynamics of a symmergent black holes spacetime are examined. We first evaluate the model under consideration's thermodynamic properties, such as its Hawking temperature, angular velocity, entropy, and electric potential. We evaluate the logarithmic correction terms for entropy around the equilibrium state in order to examine the impacts of thermal fluctuations. In the presence of these correction terms, we also examine the viability of the first law of thermodynamics. Finally, we evaluate the system's stability using the Hessian matrix and heat capacity. It is determined that a stable model is generated by logarithmic corrections arising from thermal fluctuations.
\end{abstract}

\date{\today}



\maketitle

\section{Introduction}

Hawking radiation describes hypothetical particles created near a black hole's (BHs) event horizon. By combined models of general relativity and quantum mechanics, Hawking showed that BHs are not actually black, because of the emission of a nearly thermal radiation.  This radiation implies BHs have temperatures that are inversely proportional to their mass. One can say that smaller BH is hotter. It is supposed that this radiation causes them to lose energy, shrink and eventually disappear which causes the unsolved information lost problem. There are some proposals to solve this information problem. One of them is proposed by Hawking, Perry and Strominger by defining new term in physics ‘soft particles’ \cite{Hawking1}.
There are many ways to derive Hawking radiation and calculate its temperature. The most popular one is the the semi-classical quantum tunneling strategy introduced in \cite{1}. In the tunneling phenomenon, there are two famous techniques to derive the tunneling rates. One is Parikh-Wilczek technique that consider the null geodesic equation of the radiated vector particles \cite{2}. The other way is Hamilton-Jacobi approach proposed in literature \cite{3, 4, 5, 5a, 5b, 6, 7,8,Kerner:2007rr,Kerner:2006vu,Kerner:2008qv,Kerner:2007jk,Kuang:2018goo,Kuang:2017sqa,Akhmedov:2006pg,Akhmedova:2008dz,Akhmedov:2006un,Singleton:2010gz,Kruglov:2014jya,Ali:2007sh,Ali:2007xz,Singh:2019wtn,IbungochoubaSingh:2016prd,IbungochoubaSingh:2016jkk,Singh:2017car,Meitei:2020oop}. 
The tunneling probability can be derived \cite{1a} from the following formula
\begin{equation}
\Gamma\simeq\exp\left[-\frac{2}{\hbar} Im I\right],
\end{equation}
The Hawking temperature for all types of particles can be calculated by using the above tunneling formula.
In various types of theories of quantum gravity (e.g., non-commutative geometry, loop quantum gravity and  string
theory), the main feature is the presence of a minimal noticeable length \cite{9,10}. In order to study this minimal length, the generalized Uncertainty Principle (GUP) is the most appropriate way \cite{11}. The modified commutation relationship is defined as
\begin{equation}
 [ w_{u}, p_{v}]=i\hbar\left[1+\beta(p^{2})\right]\delta_{uv},
\end{equation}
whereas $p_{v}$ and $w_{u}$ stands for generalized momentum and position operators.
Furthermore, a relation with GUP can be given in the form
\begin{equation}
    \Delta w\Delta p\geq \frac{\hbar}{2}\left[1+\beta(\Delta p^{2})\right],
\end{equation}
here the correction parameter $\beta$ can be expressed in terms of dimensionless parameter as $\beta=\frac{\beta_{0}}{M_{p^{2}}}$ and $M_{p}$ denotes the Planck's mass. The GUP association is very helpful to understand the BH physics and the quantum effects have a great influence near the horizon of a BH. Gecim and Sucu \cite{12,13,14} have studied the gravity effects incorporating GUP for $2+1$-dimensional Warped-$AdS_3$, Martinez-Zanelli and New-Type of BHs.
\"Ovg\"un et al. \cite{15,16,17} have investigated the gravity impacts via GUP in the background of tunneling method for the noncommutative Reissner-Nordstr\"om, warped DGP gravity and 5D Myers-Perry BHs and derived the temperature for corresponding BHs.
Moreover, the BH thermodynamics plays a very significant role in order to study the BH physics. The four laws of thermodynamics of BH and their relation with gravity have been investigated by Bardeen and his fellows \cite{18}. In order to study the thermodynamics of BH, one associates entropy with Bekenstein’s area whereas the temperature can be calculated through first law of thermodynamics \cite{19,20}.
Many important features of BH e.g., stable and unstable form, holographic duality, criticality and many numerous critical perspective can be studied through corrected thermodynamics of BH.
In order to study the physics of BHs, the quantum corrected fluctuations in BH thermodynamics have acquired a noteworthy place. Pourhassan et al \cite{21,21a,21b} have computed the logarithmic correction impacts of thermal fluctuations for a charged anti-de Sitter BH, Horava-Lifshitz BH and Kerr-AdS BH. Faizal and Khalil \cite{22} have investigated the GUP corrected entropy corrections for Reissner-Nordstr\"om, Kerr, charged AdS BHs as well as rotating BHs. They have also analyzed their remnant. The logarithmic corrected entropy has been investigated for Godel BH as well as by using Cardy formula \cite{23,24}.
Considering 1$^{st}$-order corrections to temperature and entropy of Kerr-Newman-anti-de Sitter BHs and Reissner-Nordstrom-anti-de Sitter BH impacts from the corrected temperature and entropy on thermodynamical quantities like enthalpy, internal energy, Gibbs
free energy and Helmholtz energy have been examined \cite{24a}. Thermal fluctuations' effects on modified Hayward BHs thermodynamics have been investigated. It has been determined \cite{24b} how these correction terms for the first law of thermodynamics will affect thermodynamic properties such as inner energy, entropy, pressure, and specific heat \cite{,Bargueno:2021nuc}.

The motivations of our study is to analyze the Hawking temperature under the impacts of quantum gravity as well as logarithmic corrections via thermodynamics of symmergent BH. Symmergent black hole is derived from a quadratic-curvature gravity which is a subclass of $f(R)$ gravity theories \cite{demir1,demir2,demir3,Symmergent-bh,Symmergent-bh2}. To compute rate of probability and the corresponding Hawking temperature, we utilize the Hamilton-Jacobi strategy
associated with generalized Proca equation for boson radiated particles. After computing the corrected temperature, we graphically discuss the stable states of symmergent BH. Moreover, by utilizing the thermal fluctuations, we study the logarithmic corrections and their graphical interpretation. 

The remaining part of this paper is organized as follows. In Section II, we briefly review the symmergent BHs. In Section III, we compute the corrected temperature of symmergent BH by utilizing the tunneling approach. Section IV, explains the physical significance and stable states of symmergent BH in the background of graphical representation of corrected temperature with horizon. In Section V, we investigates the logarithmic corrections under thermal fluctuations with the help of graphs.
We conclude our findings in Section VI.

\section{Symmergent Black hole} \label{sec2}
In general, the symmergent gravity is a gravity theory that relates with $R+R^2$ theory and it is associated with $f(R)$ gravity theories as a special case. The complete details about symmergent gravity theory can be studied in \cite{demir1,demir2,demir3}. The solution of symmergent black hole was found by {\c C}imdiker, \"Ovg\"un and Demir in \cite{Symmergent-bh} . Afterwards, the implications of this theory has been investigated in \cite{Symmergent-bh2,Pantig:2022qak} for black features e.g., weak lensing, shadow radius, quasi-periodic and oscillations.
 The spacetime metric for the symmergent black hole is given by \cite{Symmergent-bh}
\begin{equation}
ds^{2}=-f(r)dt^{2} + \frac{1}{f(r)}dr^{2}+r^{2} \left(d\theta^{2}+\sin^2\theta d\phi^{2} \right),\label{metric}
\end{equation}
with the lapse function 
\begin{equation} \label{smetric}  
f(r)=1-\frac{2 G M}{r}-\frac{1-\hat{\alpha}}{24 \pi G c_{o}} r^2,
\end{equation}

where $\hat{\alpha}$ is an integration constant and the symmergent parameter (loop coefficient) $c_{o}$ is defined as 
\begin{equation}\label{params}
c_{o} = \frac{n_{B} - n_{F}}{128 \pi^2},
\end{equation}
in which $n_{B}$/$n_{F}$ are for the total number of bosons/fermions in the underlying QFT.
The above metric function reaches to the Schwarzschild BH space-time when $\hat{\alpha}\to1$ or $c_{o}\to \infty$. The radius of event horizon for the symmergent BH is calculated by using $g^{rr}=0$. In that case we get three horizons but the largest root of event horizon can be calculated as \cite{Symmergent-bh}
\begin{equation} \label{horizons}
 r_{+}=\frac{\cal H}{\sqrt[3]{18}}-\frac{\sqrt[3]{18}a_{1}}{3\cal H}, 
\end{equation}
with 
\begin{equation}
{\cal H}=\sqrt[3]{(12a_{1}^{3}+81a_{2}^{2})^{1/2}-9a_{2}},
\end{equation} 
where $a_{1}=\frac{1}{A},~a_{2}=\frac{2G M}{A}$, $A=\frac{1-\hat{\alpha}}{24 \pi G c_{o}}$ and the BH mass $M$ can be represented in terms of the horizon, despite its complicated functional form. In order to get more details about the horizon properties check \cite{Symmergent-bh2}.

The Hawking temperature $T_H$ of a symmergent BH is calculated as follows
\begin{eqnarray}
T_H=\frac{f^{\prime}\left(r_{+}\right)}{4 \pi}=\frac{G M}{8 r_{+}^2}+\frac{(1-\hat{\alpha})r_{+}}{192\pi c_{o} G}.
\end{eqnarray}
The Bekenstein-Hawking entropy \cite{Bekenstein:1973ur,Brevik:2004sd,Symmergent-bh} is given by 
\begin{eqnarray}
S=\frac{\hat{\alpha} \pi r_{+}^2}{G}
\end{eqnarray}
which is independent of $c_{o}$. The next section analyzes the particle tunneling from the symmergent BH solution using the WKB approximation method.
\section{Particle tunnel from the symmergent black hole}

We calculate the tunneling rate of the boson particles to investigate the Hawking temperature of a symmergent BH. By using a semi-classical technique, we describe the Hawking temperature. Ali et al. \cite{r1, r2, r3, r4, r5, r6} have investigated the gravity impacts via GUP in the background of tunneling method for the BHs, cosmic strings and 5D black rings and derived the temperature for corresponding BHs.
We 
examine how quantum gravity affects the Hawking temperature in the influence of GUP. According to 
\cite{riasat1}, the gravity parameter and the BH stability feature are generally connected. The GUP 
component of the Lagrangian equation's physical importance is taken into consideration. The field 
equation without a singularity extended in the form of Lagrangian field equation is the GUP parameter. In order 
to study the boson radiation phenomena, we utilize the Lagrangian equation of action via vector field $\Psi mu$.
\begin{equation}
\partial_{\mu}(\sqrt{-g}\Psi^{\nu\mu})+\sqrt{-g}\frac{m^2}{\hslash^2}\Psi^{\nu}+
+\beta\hslash^{2}\partial_{0}\partial_{0}\partial_{0}(\sqrt{-g}g^{00}\Psi^{0\nu})
-\beta \hslash^{2}\partial_{i}\partial_{i}\partial_{i}(\sqrt{-g}g^{ii}\varphi^{i\nu})=0,\label{L}
\end{equation}
here $g$, $\Psi^{\nu\mu}$ and $m$ represents the coefficient matrix determinant, vector particle mass and anti-symmetric
tensor, respectively. Defining the anti-symmetric tensor $\Psi_{\nu\mu}$ is
\begin{equation}
\Psi_{\nu\mu}=(1-\beta{\hslash^2\partial_{\nu}^2})\partial_{\nu}\Psi_{\mu}-
(1-\beta{\hslash^2\partial_{\mu}^2})\partial_{\mu}\Psi_{\nu}\nonumber,
\end{equation}
where $\beta$ and $\hslash$ are the Plank's constant and GUP parameter, respectively.
The components of $\Psi^{\mu}$ and $\Psi^{\mu\nu}$ can be calculated as
\begin{eqnarray}
&&\Psi^{0}=-\frac{1}{f(r)}\Psi_{0},~~~\Psi^{1}=f(r)\Psi_{1},~~~\Psi^{2}=\frac{1}{r^{2}}\Psi_{2},~~~
\Psi^{3}=\frac{1}{r^{2}\sin^2\theta}\Psi_{3},~~~\Psi^{01}=-\Psi_{01},\nonumber\\
&&\Psi^{02}=-\frac{1}{f(r)r^{2}}\Psi_{02},~~~\Psi^{03}=-\frac{1}{f(r)r^{2}\sin^2\theta}\Psi_{03},~~~
\Psi^{12}=\frac{f(r)}{r^{2}}\Psi_{12},~~~\Psi^{13}=\frac{f(r)}{r^{2}\sin^2\theta}\Psi_{13},~~~
\Psi^{23}=\frac{1}{r^{4}\sin^2\theta}\Psi_{23}.\nonumber
\end{eqnarray}
The WKB strategy is given by \cite{riasat2}
\begin{equation}
\Psi_{\nu}=s_{\nu}\exp\left[\frac{i}{\hslash}T_{0}(t,r,\theta,\phi)+
\Sigma \hslash^{n}T_{n}(t,r,\theta,\phi)\right],
\end{equation}
here, $(T_{0},~T_{n})$ represents the arbitrary functions and $s_{\nu}$ is the constant term.
After neglecting the higher orders in the Lagrangian equation (\ref{L}), where the term $\hslash$ is only taken 
into account in the WKB approximation for the $1^{st}$ order, we arrive at the equation system shown below:
\begin{eqnarray}
&&f(r)\left[s_{1}(\partial_{0}T_{0})(\partial_{1}T_{0})+\beta s_{1}
(\partial_{0}T_{0})^{3}(\partial_{1}T_{0})-s_{0}(\partial_{1}T_{0})^{2}
-\beta s_{0}(\partial_{1}T_{0})^4\right]+\nonumber\\
&&\frac{1}{r^{2}}\left[s_{2}(\partial_{0}T_{0})(\partial_{2}T_{0})+\beta s_{2}
(\partial_{0}T_{0})^3(\partial_{2}T_{0})-s_{0}(\partial_{2}T_{0})^2-\beta s_{0}
(\partial_{2}T_{0})^4+\right]+\nonumber\\
&&\frac{1}{r^{2}\sin^2\theta}[s_{3}(\partial_{0}T_{0})(\partial_{3}T_{0})
+\beta s_{3}(\partial_{0}T_{0})^3(\partial_{3}T_{0})
-s_{0}(\partial_{3}T_{0})^2-\beta s_{0}(\partial_{3}T_{0})^4]-s_{0}m^2=0,\label{aa}\\
&&-\frac{1}{f(r)}\left[s_{0}(\partial_{0}T_{0})(\partial_{1}T_{0})+\beta s_{0}
(\partial_{0}T_{0})(\partial_{1}T_{0})^3-s_{1}(\partial_{0}T_{0})^{2}
-\beta s_{1}(\partial_{0}T_{0})^{4}\right]+\nonumber\\
&&\frac{1}{r^{2}}\left[s_{2}(\partial_{1}T_{0})(\partial_{2}T_{0})+\beta s_{2}
(\partial_{1}T_{0})^3(\partial_{2}T_{0})-s_{1}(\partial_{2}T_{0})^{2}-\beta
s_{1}(\partial_{2}T_{0})^{4}\right]+\nonumber\\&&
\frac{1}{r^{2}\sin^2\theta}\left[s_{3}(\partial_{1}T_{0})(\partial_{3}T_{0})+\beta s_{3}
(\partial_{1}T_{0})^3(\partial_{3}T_{0})
-s_{1}(\partial_{3}T_{0})^{2}-\beta s_{1}(\partial_{3}T_{0})^{4}\right]-s_{1}m^2=0,\\
&&-{\frac{1}{f(r)}}\left[s_{0}(\partial_{0}T_{0})(\partial_{2}T_{0})+\beta s_{0}
(\partial_{0}T_{0})(\partial_{2}T_{0})^{3}-s_{2}(\partial_{0}T_{0})^{2}
-\beta s_{2}(\partial_{0}T_{0})^4\right]\nonumber\\
&&+f(r)\left[s_{1}(\partial_{1}T_{0})(\partial_{2}T_{0})+\beta s_{1}
(\partial_{1}T_{0})(\partial_{2}T_{0})^{3}
-s_{2}(\partial_{1}T_{0})^{2}-\beta s_{2}(\partial_{1}T_{0})^4\right]+\nonumber\\
&&\frac{1}{r^{2}\sin^2\theta}\left[s_{3}(\partial_{2}T_{0})(\partial_{3}T_{0})+\beta s_{3}
(\partial_{2}T_{0})^{3}(\partial_{3}T_{0})-s_{2}(\partial_{3}T_{0})^{2}-\beta s_{2}(\partial_{3}T_{0})^4\right]+m^2 s_{2}=0,\\
&&-{\frac{1}{f(r)}}\left[s_{0}(\partial_{0}T_{0})(\partial_{3}T_{0})+\beta s_{0}
(\partial_{0}T_{0})(\partial_{3}T_{0})^{3}-s_{0}(\partial_{3}T_{0})^{2}-
s_{0}(\partial_{3}T_{0})^{4}\right]-\nonumber\\
&&f(r)\left[s_{1}(\partial_{1}T_{0})(\partial_{3}T_{0})+\beta s_{1}
(\partial_{1}T_{0})(\partial_{3}T_{0})^{3}-s_{3}(\partial_{1}T_{0})^{2}
-\beta s_{3}(\partial_{1}T_{0})^4\right]+\nonumber\\
&&\frac{1}{r^{2}}\left[s_{2}(\partial_{2}T_{0})(\partial_{3}T_{0})+\beta s_{2}
(\partial_{2}T_{0}){(\partial_{3}T_{0})^3}-s_{3}(\partial_{2}T_{0})^{2}-\beta s_{3}(\partial_{2}T_{0})^4\right]-m^2 s_{3}=0.\label{ab}
\end{eqnarray}
We take into consideration the idea of variable separation
\begin{equation}
T_{0}=-\hat{E}t+B(r,\theta)+J\phi,\label{c1}
\end{equation}
where $\hat{E} =(E-J\Omega)$ with $J$ and $E$ indicate the particle angular momentum and energy of particle at angle $\phi$, respectively.
We obtain a matrix of order $4\times 4$ in the following way by applying the Eq. (\ref{c1}) into Eqs. (\ref{aa})-(\ref{ab}).

\begin{equation*}
A(s_{0},s_{1},s_{2},s_{3})^{T}=0.
\end{equation*}
The specified matrix appears to be non-trivial. The components of it are mentioned below:
\begin{eqnarray}
A_{00}&=&-B_{r}^2-\beta B_{r}^4-\frac{1}{f(r)r^{2}}(B_{\theta}+\beta B_{\theta}^4)
-\frac{1}{f(r)r^{2}\sin^2\theta}(\dot{J}^2+\beta\dot{J}^4)-\frac{1}{f(r)}m^2,\nonumber\\
A_{01}&=&-(\hat{E}+\beta \hat{E}^3)B_{r},\nonumber\\
A_{02}&=&-\frac{1}{f(r)r^{2}}(\hat{E}+\beta \hat{E}^3)B_{\theta},\nonumber\\
A_{03}&=&-\frac{1}{f(r)r^{2}}(\hat{E}+\beta \hat{E}^3)\dot{J},\nonumber\\
A_{10}&=&\hat{E}B_{r}+\beta \hat{E}^3 B_{r}^4,\nonumber\\
A_{11}&=&\hat{E}^2+\beta \hat{E}^4-
\frac{f(r)}{r^{2}}(B_{\theta}^2+\beta B_{\theta}^4)
-\frac{f(r)}{r^{2}\sin^2\theta}(\dot{J}^2+\beta\dot{J}^4)-f(r)m^2,\nonumber\\
A_{12}&=&\frac{f(r)}{r^{2}}(B_{r}+\beta B_{r}^3)B_{\theta},~~
A_{13}=\frac{f(r)}{r^{2}\sin^2\theta}(B_{r}+\beta B_{r}^3)B_{\theta},\nonumber\\
A_{20}&=&-\frac{1}{f(r)r^{2}}(\hat{E}B_{\theta}+\beta \hat{E}B_{\theta}^3)\nonumber\\
A_{21}&=&\frac{f(r)}{r^{2}}(B_{\theta}+\beta B_{\theta}^3)B_{r},\nonumber\\
A_{22}&=&\frac{1}{f(r)r^{2}}(\hat{E}^2+\beta \hat{E}^4)
-\frac{f(r)}{r^{2}}(B_{r}^2+\beta B_{r}^4)-\frac{1}{r^{4}\sin^2\theta}(\dot{J}^2+\beta\dot{J}^4)
-\frac{1}{r^{2}}m^2\nonumber\\
A_{23}&=&\frac{1}{r^{4}\sin^2\theta}(B_{r}+\beta B_{r}^3)B_{\theta},\nonumber\\
A_{30}&=&\hat{E}(\dot{J}+\beta \dot{J}^3),\nonumber\\
A_{31}&=-&\frac{f(r)}{r^{2}\sin^2\theta}(\dot{J}+\beta \dot{J}^3)B_{r},~~
A_{32}=\frac{1}{r^{4}\sin^2\theta}(\dot{J}+\beta \dot{J}^3)B_{\theta},\nonumber\\
A_{33}&=&\hat{E}^2+\beta\hat{E}^4+
\frac{f(r)}{r^{2}\sin^2\theta}(B_{r}^2+\beta B_{r}^4)-\frac{1}{r^{4}\sin^2\theta}(B_{\theta}^2+\beta B_{\theta}^4)-
\frac{1}{r^{2}\sin^2\theta}m^2,\nonumber
\end{eqnarray}
where $\dot{J}=\partial_{\phi}T_{0}$, $B_{r}=\partial_{r}{T_{0}}$ and $B_{\theta}=\partial_{\theta}{T_{0}}$.
Considering that the determinant of A is a non-trivial matrix result, set A to zero, which causes the 
imaginary part to act in the form:
\begin{equation}\label{a1}
ImB^{+}=+ \int\sqrt{\frac{\hat{E}^{2}
+X_{1}\left[1+\beta\frac{X_{2}}{X_{1}}\right]}{f(r)r^{-2}\sin^{-2}\theta}^2}dr,
\end{equation}
with
\begin{equation}
X_{1}=-\frac{B_{\theta}}{r^{2}\sin^{2}\theta}-\frac{m^2}{r^{2}\sin^{2}\theta},~~~
X_{2}=\frac{f(r)}{r^{2}\sin^{2}\theta}B_{r}^4+\hat{E}^{4}+\frac{B_{r}^4}{r^{2}\sin^{2}\theta}.\nonumber
\end{equation}
The Eq. (\ref{a1}) implies
\begin{equation}
Im B^{+}=\pi\frac{\hat{E}}{2\kappa(r_{+})}\left[1+\beta \Xi\right],
\end{equation}
here $\Xi$ indicates the arbitrary parameter.
The modified tunneling rate for boson particles can be determined by using the formula:
\begin{equation}
T(B^{+})=\exp\left[-4Im B^+\right]=
\exp\left[{-2\pi}\frac{\hat{E}}
{\kappa(r_{+})}\right]\left[1+\Xi\beta\right],
\end{equation}
where
\begin{equation}
\kappa(r_{+})=\frac{1}{2}\frac{1}{\sqrt{-g_{tt}g_{rr}}}\left|g_{tt,r}\right|_{r=r_{+}}.
\end{equation}
The Hawking temperature for symmergent BH under the effect of GUP parameter can
be derived by utilizing Boltzmann factor $T_{B}=\exp\left[-\hat{E}/T' _{H}\right]$ as follows:
\begin{equation}
T'_{H}=\left(\frac{G M}{8 r_{+}^2}-\frac{(1-\hat{\alpha})r_{+}}{192 \pi G c_\text{O}} \right)\left[1-\beta\Xi\right].\label{F5}
\end{equation}
As we can see, the quantum corrections and the BH geometry both have an impact on the corrected Hawking 
temperature. The zero order correction term is the same as the semi-classical original Hawking term, but 
the first-order correction term must be smaller than the preceding term while still satisfying GUP. The $T'_{H}$ depends on the BH mass $M$, arbitrary parameter $\Xi, \hat{\alpha}$, symmergent gravity parameter $G$, loop parameter $c_{o}$ and correction parameter $\beta$. Moreover, after neglecting the gravity parameter $\beta=0$ into equation (\ref{F5}), we recover the original temperature of symmergent BH in \cite{Symmergent-bh}.

\section{Graphical Analysis of $T'_{H}$ for symmergent BH}
This section analyze the effects of gravity parameter and $\hat{\alpha}$ on $T'_{H}$ for symmergent BH.
We investigate the stable condition of symmergent BH under the effects of quantum gravity parameter
by fixing the values of arbitrary parameter and symmergent gravity parameter $\Xi=1=G$ in the regions $0\leq M\leq10$ and $-1\leq c_{o}\leq1$.

\begin{center}
\includegraphics[width=8.4cm]{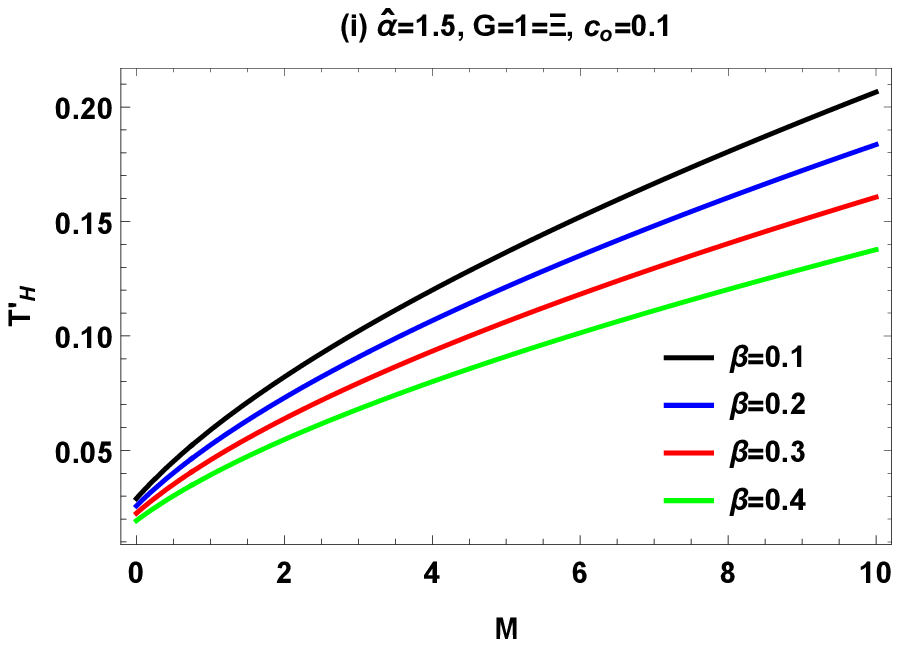}\includegraphics[width=8.4cm]{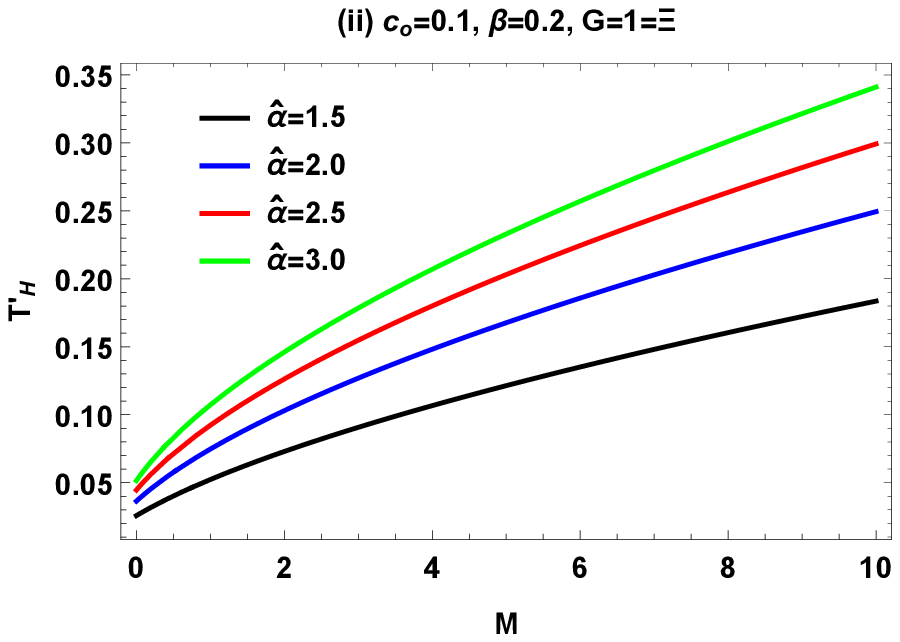}\\
{Figure 1: $T'_{H}$ versus $M$} for fixed  $\Xi=1=G$.
\end{center}

\textbf{Figure 1}: (i) shows the graphical interpretation of $T'_{H}$ with Mass $M$ in the domain $0\leq M\leq 10$ for constant values of $c_{o}=0.1$, $\hat{a}=1.5$ as well as for different values of correction parameter $\beta$.
One can observe that the temperature exponentially increases with varying mass. The $T'_{H}$ decreases as we increase the values of correction parameter $\beta$. So, we can say the quantum corrections cause a reduction in the rise of temperature.

(ii) analyze the conduct of $T'_{H}$ versus $M$ for various values of arbitrary parameter $\hat{a}$ and constant values of $\beta=0.2, c_{o}=0.1$. It can be seen that, the $T'_{H}$ increases with increasing values of $M$ with positive temperature. This positive conduct of temperature states the stable condition of BH. The $T'_{H}$ increases with the increasing values of parameter $\hat{\alpha}$.
\begin{center}
\includegraphics[width=8.4cm]{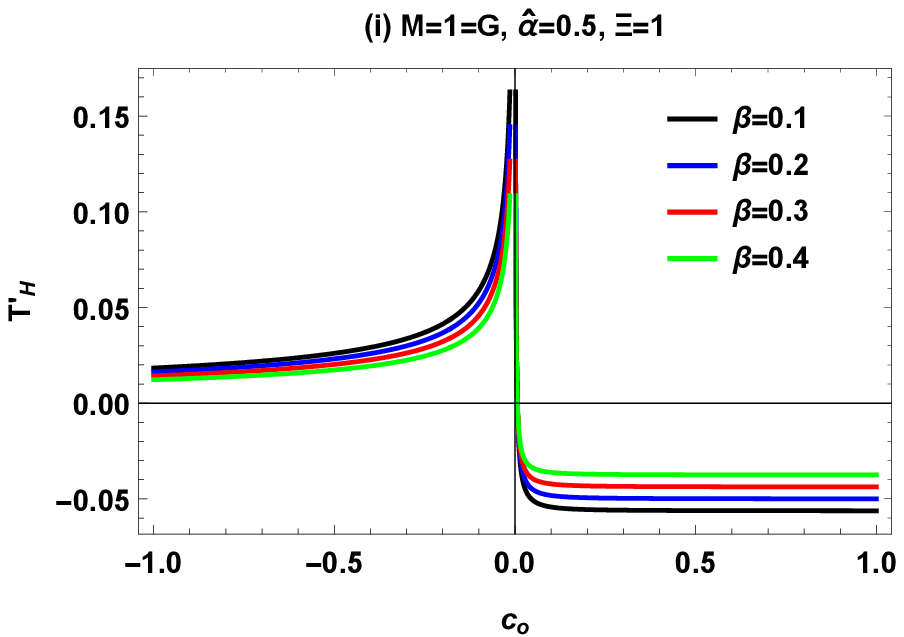}\includegraphics[width=8.4cm]{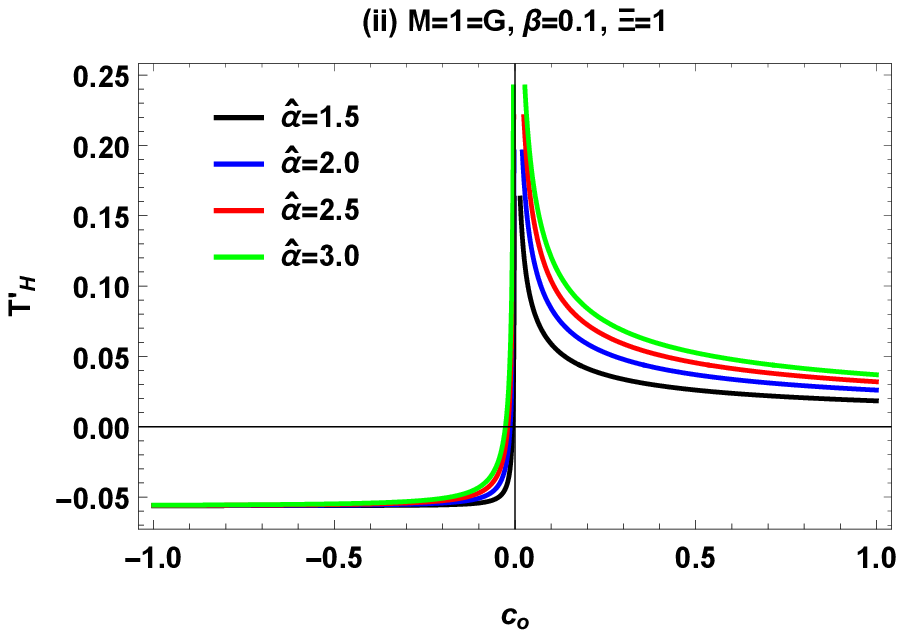}\\
{Figure 2: $T'_{H}$ versus $c_{o}$} for fixed $M=1=G$ and $\Xi=1$.
\end{center}
\textbf{Figure 2}: (i) depicts the graphical analysis of $T'_{H}$ via $c_{o}$ in the range $-1\leq c_{o}\leq 1$ for constant values of $\hat{a}=0.5$ and changing values of correction parameter $\beta$.
For $c_{o}<0$, the region shows the behavior of temperature when total number of fermions are larger than a total number of bosons while for $c_{o}>0$ shows the behavior of temperature when when total number of bosons are larger than a total number of fermions. It is also observable that the $T'_{H}$ decreases for increasing values of $\beta$ in the region $c_{o}<0$ whereas it increases for increasing values of $\beta$ in the region $c_{o}>0$.

(ii) describes the conduct of $T'_{H}$ versus $c_{o}$ for different values of $\hat{\alpha}$ and constant values of $\beta=0.1$. It can be observed that for $\hat{\alpha}>1$ and $c_{O}<0$, the temperature shows negative behavior while for  $\hat{\alpha}>1$ and $c_{O}>0$, the temperature shows positive decreasing behavior. This type of conduct depicts the physical and stable form of BH.
It has also worth mentioning here that, a deceleration in temperature can be observed for negative values of $c_{o}<0$ as compared to positive values of $c_{o}>0$.

\section{Thermal fluctuations}

In order to explore the impact of thermal fluctuations \cite{re1}, several thermodynamical potentials of the rotating BTZ BHs have been determined, and the Hawking temperature and corrected entropy are known. In this regard, the system's leading-order corrected enthalpy energy for small BHs takes on an asymptotic value that corresponds to the correction parameter. They discovered a critical threshold below which the effects of thermal fluctuation are negligible and the logarithmic entropy correction for BTZ like BH, hairy BHs, Schwarzschild BH and Reissner-Nordstrom BH have been studied \cite{re2, re3, re4}.
The study of BH thermodynamics is greatly influenced by thermal fluctuations. The idea of Euclidean quantum gravity causes a shift in the temporal coordinates in favor of complex plans. The partition function $Z(\mu)$ in terms of density of states $\eta(E)$ is provided \cite{ZR} as a way to verify the corrected entropy along these thermal fluctuations \cite{Gibbons:1976ue,Iyer:1995kg}
\begin{equation}
Z(\xi)=\int_{0}^{\infty} \exp(-\xi E)\eta(E)dE,
\end{equation}
where $T_{+}=\frac{1}{\xi}$ and E is the thermal radiation's average energy. The equation of density forms under the Laplace inverse transform
\begin{equation}
\rho(E)=\frac{1}{2\pi
i}\int_{\xi_{0}-i\infty}^{\xi_{0}+i\infty} Z(\xi)
\exp(\xi E) d\xi=\frac{1}{2\pi
i}\int_{\xi_{0}-i\infty}^{\xi_{0}+i\infty}
\exp(\tilde{S}(\xi))d\xi,
\end{equation}
where $\tilde{S}(\xi)=\xi E+ \ln Z(\xi)$ is known as the corrected entropy of the considered system. Under the steepest decent method, the equation of corrected entropy takes the the following form \cite{Das:2001ic,Sadeghi:2014zna}
\begin{equation}
\tilde{S}(\xi)=S+\frac{1}{2}(\xi-\xi_{0})^{2}
\frac{\partial^{2}\tilde{S}(\xi)}{\partial
\xi^{2}}\Big|_{\xi=\xi_{0}}+\text{higher-order terms}.
\end{equation}
By using the conditions $\frac{\partial
\tilde{S}}{\partial\xi}=0$ and $\frac{\partial^{2}
\tilde{S}}{\partial\xi^{2}}>0$, the corrected entropy relation under the first-order corrections has been modified \cite{ZR} and by neglecting the higher order terms, the exact expression of entropy is expressed as \cite{21}
\begin{equation}\label{c}
\tilde{S}=S-\psi \ln(ST^{2}),
\end{equation}
where $\psi$ represents the logarithmic correction parameter. Pourhassan and Faizal have analyzed the effects of thermal fluctuations on a charged AdS black hole by introducing, for the first time, the corrected form of entropy in literature \cite{21}. Then  using the Bekenstein-entropy and Hawking temperature into Eq. (\ref{c}), we get
\begin{equation}
\tilde{S}=\frac{\pi  \hat{\alpha}  r_+^2}{G}+\psi  \left(\log (36864 \pi )-\log \left(\frac{\hat{\alpha}  \left((\hat{\alpha} -1)
   r_+^3-24 \pi  G^2 M c_{o}\right){}^2}{G^3 r_+^2
   c_{o}^2}\right)\right)
   \end{equation}.

\begin{center}
\includegraphics[width=8cm]{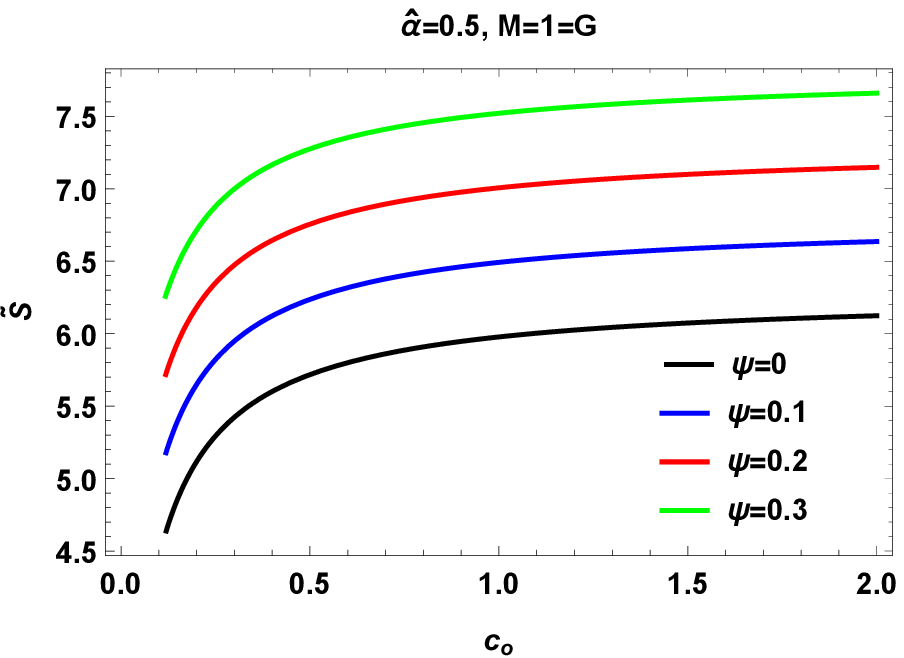}\\
{Figure 3: Corrected entropy with respect to $c_{o}$ for fixed $M=1=G$ and $\hat{\alpha}=0.5$ and varying $\psi$}.
\\
\end{center}
From \textbf{Fig. 3}, the behaviour of corrected entropy along $c_{o}$ is monotonically increasing. It is noted that for the graph of usual entropy shows increasing behaviour for small values of $c_{o}$, so these logarithmic corrections are more useful for small BHs. With the help of corrected entropy, we can also check behaviour of other thermodynamic quantities in the influence of these corrections. So, the Helmholtz energy ($F=-\int \tilde{S}dT$) leads to the form
\begin{eqnarray}
F&=&\frac{(\hat{\alpha} -1) \hat{\alpha}  r_+^3}{576 G^2 c_{o}}+\frac{1}{4} \pi  \hat{\alpha}
   M \psi \log \left(r_+\right).
\end{eqnarray}
 \begin{center}
\includegraphics[width=8cm]{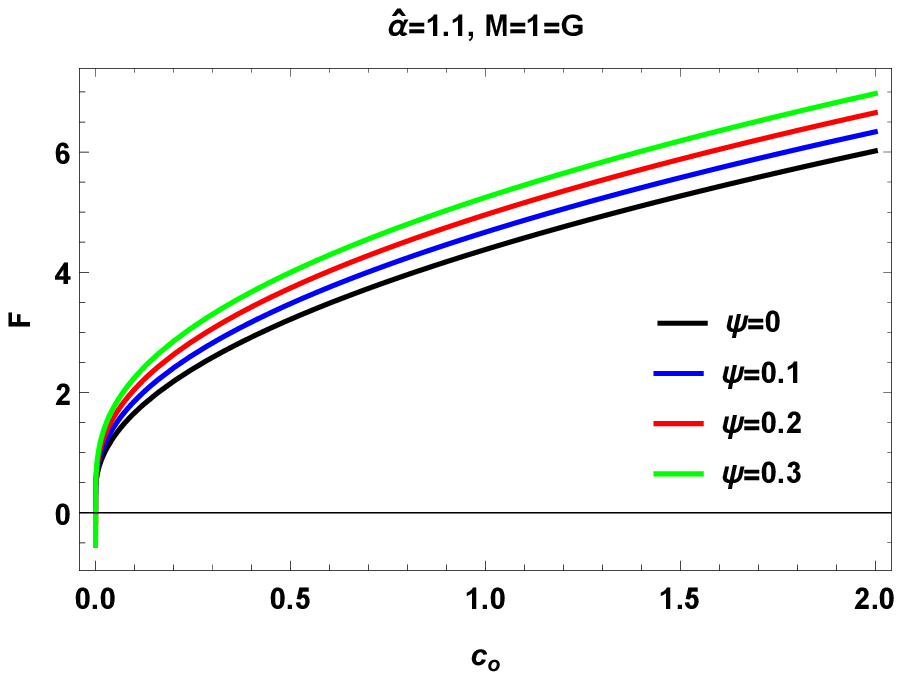}\\
{Figure 4: Helmholtz free energy with respect to $c_{o}$ for fixed $M=1=G$, $\hat{\alpha}=1.1$ and varying $\psi$}.
\\
\end{center}
\textbf{Fig. 4} represents the curve of Helmholtz free energy along $c_{o}$. For the different choices of correction parameter, the Helmholtz free energy depicts increasing behaviour, it means that the system under consideration changes state to equilibrium. Moreover, there is another important thermodynamic quantity internal energy ($E=F+T\tilde{S}$) is given as
\begin{eqnarray}
E&=&\frac{(\hat{\alpha} -1) \hat{\alpha}  r_+^3}{576 G^2 c_{o}}+\Big(\frac{G M}{8
   r_+^2}-\frac{(\hat{\alpha} -1) r_+}{192 \pi  G c_{o}}\Big) \Big(\psi 
   \Big(\log (36864 \pi )-\log \Big(\frac{\hat{\alpha}  \left((\hat{\alpha} -1)
   r_+^3-24 \pi  G^2 M c_{o}\right){}^2}{G^3 r_+^2
   c_{o}^2}\Big)\Big)\nonumber\\&+&\frac{\pi  \hat{\alpha}  r_+^2}{G}T\big)+\frac{1}{4}
   \pi \hat{\alpha}  M \log \left(r_+\right).
\end{eqnarray}
\begin{center}
\includegraphics[width=8cm]{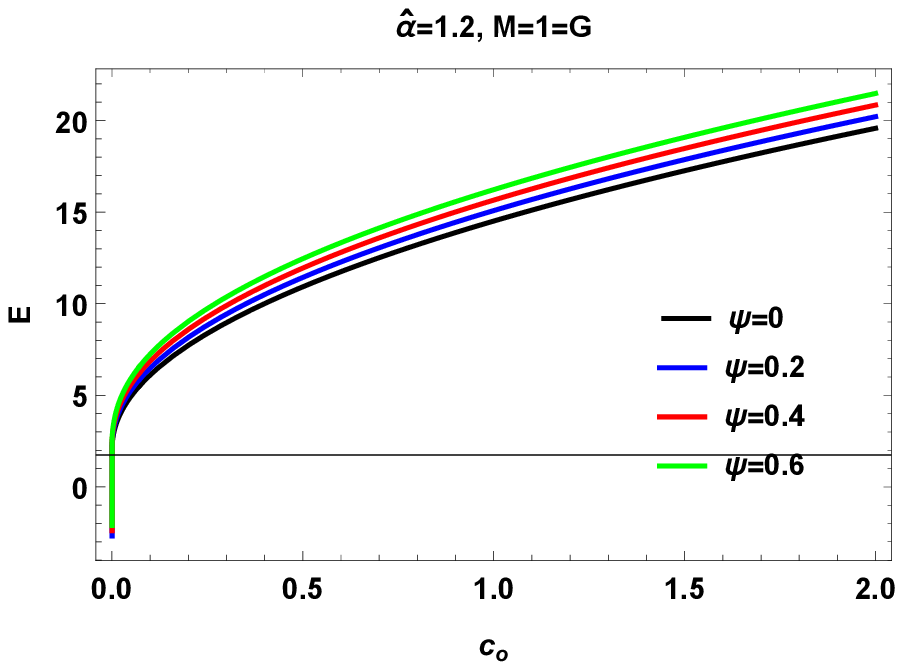}\\
{Figure 5: Internal energy with respect to $c_{o}$ for $\hat{a}=1.2$ and $M=1=G$}.
\\
\end{center}
In\textbf{Fig. 5}, the curves of internal energy depicts gradually increasing behaviour for the small values of $c_{o}$. The corrected internal energy shows increasing behaviour (To sustain its state, the BH must absorb increasing amounts of heat from the environment). Furthermore, Pressure is a further significant thermodynamic quantity. The expression of BH pressure ($P=-\frac{dF}{dV}$) for these corrections is given as
\begin{eqnarray}
P&=& \frac{\hat{\alpha}  \Big(\frac{(\hat{\alpha} -1) r_+^3}{G^2 c_{o}}+48 \pi 
   M\psi\Big)}{192 r_+}.
\end{eqnarray}
 \begin{center}
\includegraphics[width=8cm]{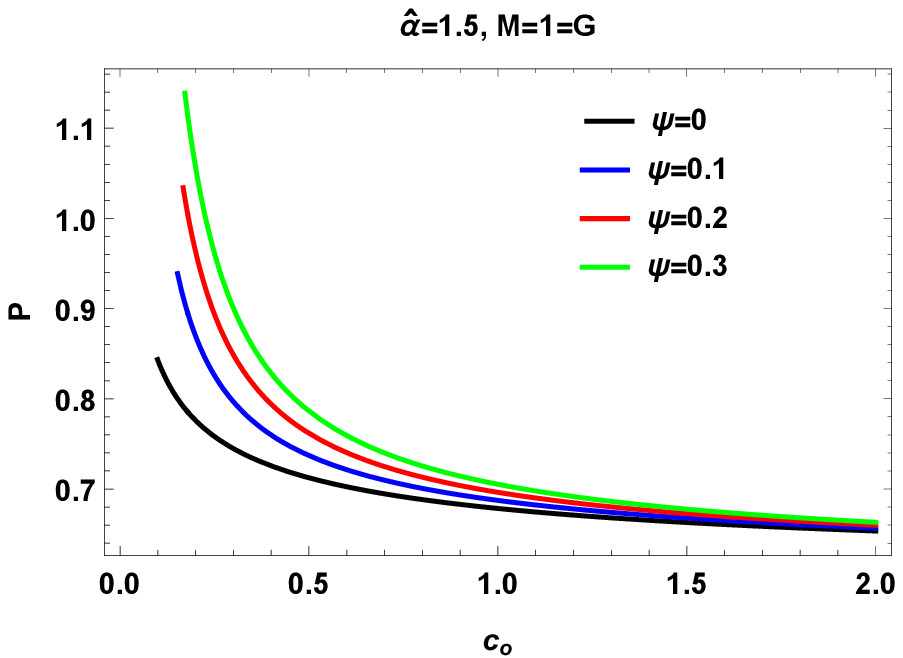}\\
{Figure 6: Pressure with respect to $c_{o}$ for $\hat{a}=1.5$ and $M=1=G$}
\\
\end{center}
\textbf{Fig. 6} shows how closely the pressure graph and the equilibrium state correspond. For various correction parameter values, the pressure dramatically reduces for the considered geometry. Enthalpy ($H=E+PV$) is another significant thermodynamic quantity, is described as
\begin{eqnarray}
H&=& \frac{\hat{\alpha} \Big(\frac{(\hat{\alpha} -1) r_+^3}{G^2 c_{o}}+48 \pi 
   M\psi\Big)}{192 r_+}+\frac{(\hat{\alpha} -1) \hat{\alpha}  r_+^3}{576 G^2
   c_{o}}+\Big(\frac{G M}{8 r_+^2}-\frac{(\hat{\alpha} -1) r_+}{192 \pi  G
   c_{o}}\Big) \Big(\psi  \Big(\log (36864 \pi )-\log
   \Big(\frac{\hat{\alpha}  \Big((\hat{\alpha} -1) r_+^3-24 \pi  G^2 M
   c_{o}\Big){}^2}{G^3 r_+^2 c_{o}^2}\Big)\Big)\nonumber\\&+&\frac{\pi  \hat{\alpha} 
   r_+^2}{G}\Big)+\frac{1}{4} \pi  \hat{\alpha}  M \log \Big(r_+\Big).
\end{eqnarray}
 \begin{center}
\includegraphics[width=8cm]{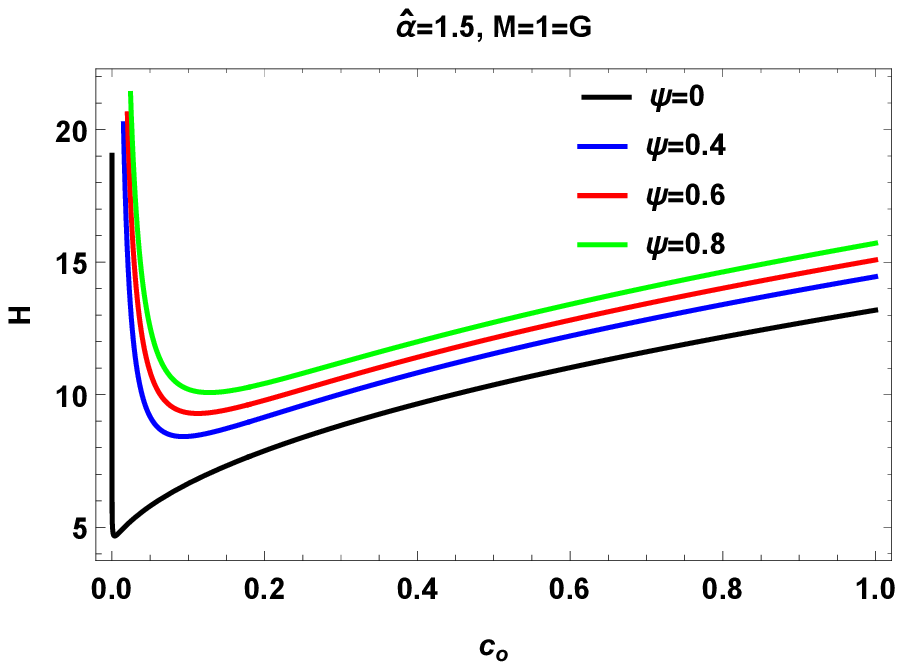}\\
{Figure 7: Enthalpy with respect to $c_{o}$ for $\hat{a}=1.5$ and $M=1=G$}.
\\
\end{center}
\textbf{Fig. 7} shows that the usual enthalpy graph eventually downs and then exponentially increases. This implies that exothermic reactions exist and that a significant amount of energy will be released into the environment. Under the effect of thermal fluctuations, the Gibbs free energy ($\tilde{G}=H-T\tilde{S}$) is expressed as
\begin{eqnarray}
\tilde{G}&=&\frac{( \hat{\alpha} -1) \hat{\alpha}  \left(r_++3\right) r_+^2}{576 G^2
   c_{o}}+\frac{\pi   \hat{\alpha}  M \psi \left(r_+ \log \left(r_+\right)+1\right)}{4
   r_+}.
\end{eqnarray}
\begin{center}
\includegraphics[width=8cm]{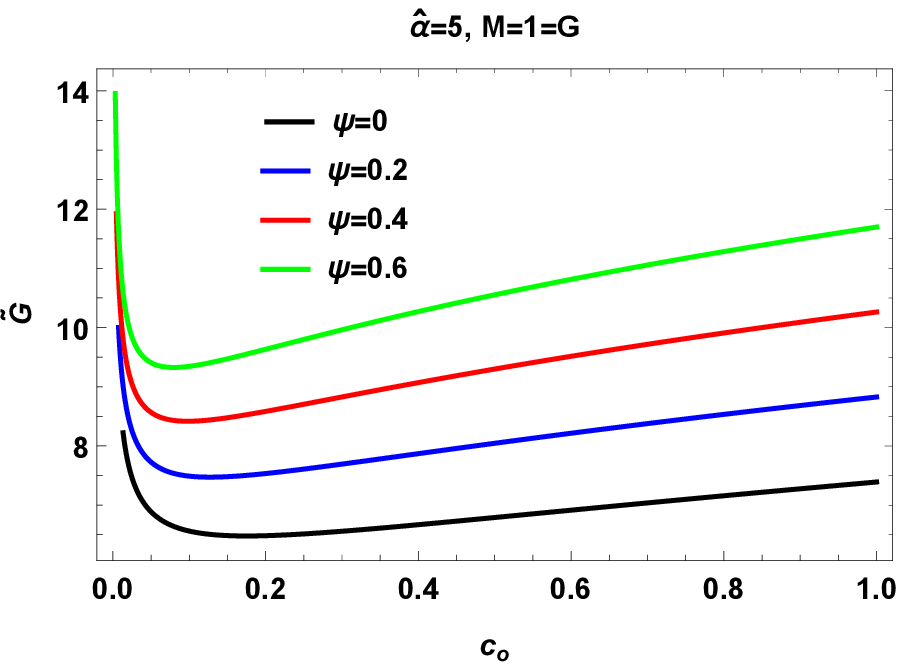}\\
{Figure 8: Gibbs free energy with respect to $c_{o}$ for $\hat{a}=5$ and $M=1=G$}.
\\
\end{center}
 \textbf{Fig. 8} illustrates the graphical analysis of the Gibbs free energy with respect to $c_{o}$. Positive energy indicates the presence of non-spontaneous processes, which means this system needs more energy to reach equilibrium. After a thorough analysis of thermodynamic quantities, another crucial idea is the system's stability as determined by specific heat. The specific heat  ($C_{\tilde{S}}=\frac{dE}{dT}$) is given as
\begin{eqnarray}
C_{\tilde{S}}&=&\Big((\hat{\alpha} -1) r_+^3 \Big(G \psi  \Big(-\log \Big(\frac{  \hat{\alpha}
   \Big((  \hat{\alpha} -1) r_+^3-24 \pi  G^2 M c_{o}\Big){}^2}{G^3 r_+^2
   c_{o}^2}\Big)-4+\log (\pi )+\log (36864)\Big)+2 \pi    \hat{\alpha}
   r_+^2\Big)\nonumber\\&-&48 \pi  G^2 M c_{o} \Big(G \psi  \Big(\log
   \Big(\frac{  \hat{\alpha}  \Big((  \hat{\alpha} -1) r_+^3-24 \pi  G^2 M
   c_{o}\Big){}^2}{G^3 r_+^2 c_{o}^2}\Big)+1-\log (36864 \pi
   )\Big)+\pi   \hat{\alpha}  r_+^2\Big)\Big)\Big(48 \pi  G^3 M c_{o}\nonumber\\&+&(  \hat{\alpha} -1) G
   r_+^3\Big)^{-1}.
\end{eqnarray}
 \begin{center}
  \includegraphics[width=8cm]{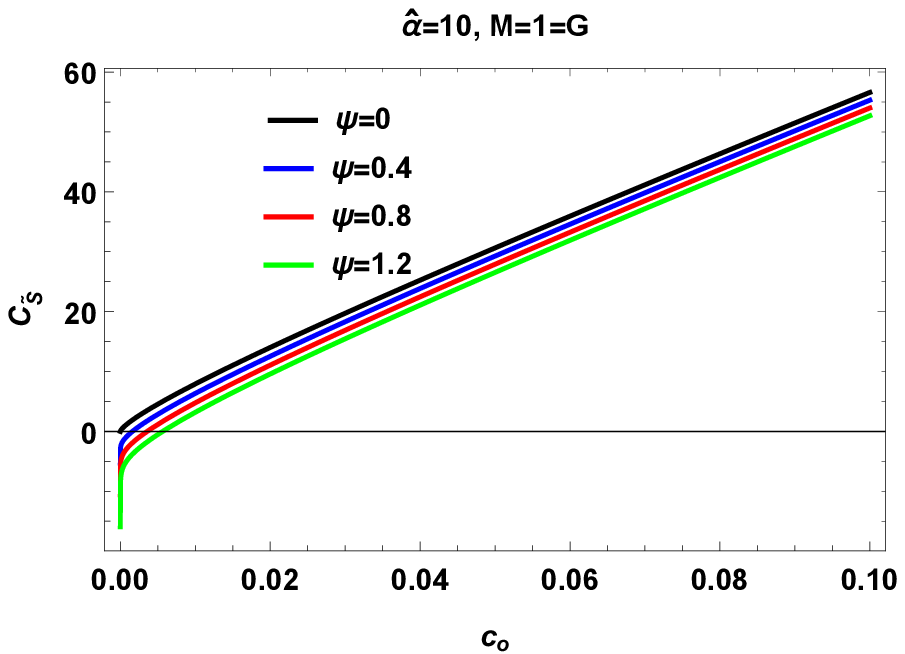}\\
{Figure 9: Specific heat with respect to $c_{o}$ for $\hat{a}=10$ and $M=1=G$}.
\\
\end{center}
\textbf{Fig. 9} shows the behavior of specific heat with respect to $c_{o}$. It is clear that whereas the uncorrected quantity (black) represents the specific heat lower than zero, indicating that the system is unstable, the corrected specific heat exhibits positive behavior across the whole examined area. This plot's positivity identifies the region of stability. It is evident that the system is stable under the influence of these corrections.
\section{Summary and Discussion}
In this article, we have computed the corrected Hawking temperature for symmergent BH under the influence of GUP parameter. At first, we have studied about the metric of an exact spherically symmetric BH in the background of  an symmergent gravity known as symmergent gravity. The gravity theory develops from
quantum loops of the fundamental quantum field theory with the symmergent gravity constant $G$ and the loop parameter $c_{o}$. In order to study the quantum corrected temperature for symmergent BH, 
we have used the semi-classical approach, WKB approximation and modified wave equation for spin-$1$ particles in the background of GUP. After accompanying the WKB approximation into modified wave equation, we have attained a set of field equations, and by considering the separation of variables strategy, we have derived a matrix of order four by four and after putting the determinant of the matrix equals to zero, we have calculated a non-trivial solution as an imaginary part of the particle action of bosons. We have investigated the tunneling probability and modified temperature for the symmergent BH at horizon $r_{+}$ by using the Boltzmann factor $T_{B}=\exp\left[\hat{E}/{T'_{H}}\right]$. 
It has worth to mention here that the both self-gravitational and back-reaction effects of the spin-$1$ particles on this symmergent BH have been neglected and the $T'_{H}$ have been computed as a leading term.
The $T'_{H}$ depends on the BH mass $M$, arbitrary parameter $\Xi, \hat{\alpha}$, symmergent gravity parameter $G$, loop parameter $c_{o}$ and correction parameter $\beta$. Moreover, after neglecting the gravity parameter $\beta=0$ into equation (\ref{F5}), we recover the original temperature for symmergent BH in \cite{Symmergent-bh}.

In order to analyze the gravity effects on the symmergent BH, we have studied the graphical analysis of the corrected Hawking temperature versus mass $M$ and loop parameter $c_{O}$ under the effects of various parameters. We have graphically investigated the physical state and stability of BH under the influence of arbitrary parameter $\hat{\alpha}$, and quantum gravity parameter $\beta$. 
It can be observed that the temperature exponentially increases with varying mass in the region $0\leq M\leq 10$. The $T'_{H}$ decreases with increasing values of correction parameter $\beta$. So, we conclude that quantum corrections cause a reduction in the rise of temperature.
Moreover, we have analyzed the conduct of $T'_{H}$ versus $M$ for various values of arbitrary parameter $\hat{a}$. The $T'_{H}$ increases with increasing values of $M$ with positive temperature. The positive conduct of temperature states the stable condition of BH. The $T'_{H}$ increases with the increasing values of parameter $\hat{\alpha}$.

We have depicted the graphical analysis of $T'_{H}$ via $c_{o}$ in the range $-1\leq c_{o}\leq 1$ for changing values of correction parameter $\beta$.
For $c_{o}<0$, we have observed the behavior of temperature when total number of fermions are larger than a total number of bosons while for $c_{o}>0$, we have observed the behavior of temperature when the total number of bosons are larger than a total number of fermions. It is also observed that the $T'_{H}$ decreases for increasing values of $\beta$ in the region $c_{o}<0$ whereas it increases for increasing values of $\beta$ in the region $c_{o}>0$.

Furthermore, we have analyzed the conduct of $T'_{H}$ versus $c_{o}$ for different values of $\hat{\alpha}$. It can be observed that for $\hat{\alpha}>1$ and $c_{O}<0$, the temperature shows negative behavior while for  $\hat{\alpha}>1$ and $c_{O}>0$, the temperature shows positive decreasing behavior. This type of conduct depicts the physical and stable form of BH.
It has also worth to mention here that, a deceleration in temperature can be observed for negative values of $c_{o}<0$ as compared to positive values of $c_{o}>0$.

\begin{acknowledgements}A. {\"O}.
 would like to acknowledge the contribution of the COST Action CA18108 - Quantum gravity phenomenology in the multi-messenger approach (QG-MM).
\end{acknowledgements}

\bibliography{ref}
\end{document}